# Beam Space Propulsion*

**Alexander Bolonkin**
C&R, 1310 Avenue R, #F-6, Brooklyn, NY 11229, USA
T/F 718-339-4563, aBolonkin@juno.com, http://Bolonkin.narod.ru

## Abstract

Author offers a revolutionary method - non-rocket transfer of energy and thrust into Space with distance of millions kilometers. The author has developed theory and made the computations. The method is more efficient than transmission of energy by high-frequency waves. The method may be used for space launch and for acceleration the spaceship and probes for very high speeds, up to relativistic speed by current technology. Research also contains prospective projects which illustrate the possibilities of the suggested method.

*Presented to Conference "Space-2006", 19-21 September, 2006, San-Jose, CA, USA.

## Introduction

Transportation of energy, matter, or impulse is very important for long period space trips especially for lengthy distance voyages. The spaceship crew or astronauts on planets can need additional energy or ship thrust. Most people think that is impossible to transfer energy a long distance in outer space except electromagnetic waves. Unfortunately, electromagnetic waves have a big divergence and cannot be used at a long distance (millions of kilometers) transfer.

However, the space vacuum is very good medium for offered method and special transfer of energy and momentum.

**Brief history**. About 40 years ago scientists received plasma flow having speed up 1000 km/s, power 10 kW, mass consumption 0.1 g/s, electric current up million amperes.

However, the application of plasma beam into space needs a series of inventions, innovations and researches. In particular, they include methods of decreasing the plasma divergence, discharging, dispersion of velocity, collection the plasma beam in space at long distance from source, conversion of the beam energy into electricity and other types of energy, conversion of plasma impulse (momentum) in space apparatus thrust, conversion of plasma into matter, control, etc.

The author started this research more then forty years ago [1]. The solutions of the main noted problems and innovations are suggested by author in early (1982-1983) patent applications [2] - [12] (see also further development in [13]-[31]) and given article. In particularly, the main innovations are:

1. Using neutral plasma (not charged beam);
2. Using ultra-cool plasma or particle beam in conventional temperature;
3. Control electrostatic collector which separates and collects the ions at spaceship;
4. Control electrostatic generator which convert the ion kinetic energy into electricity;
5. Control electrostatic ramjet propulsion;
6. Special control electrostatic mirror-reflector;
7. Recombination photon engine;
8. Recombination thermo-reactor.

Research is made for conventional and relativistic particle speeds.

About 20 years ago the scientists received the ultra-cold plasma having the ion temperature lower then $1 \times 10^{-3}$ °K. Velocity dispersion was $10^{-4} \div 10^{-6}$, beam divergence for conventional temperature was $10^{-3}$ radian.

If plasma accelerator is designed special for getting the ultra-cold plasma, its temperature may be appreciably decreased. There is no big problem in getting of cold ions from solid electrodes or cold electrons from solid points where molecular speed is small.

## Description of innovation

Innovative installation for transfer energy and impulse includes (fig. 1): the ultra-cold plasma injector, electrostatic collector, electrostatic electro-generator-thruster-reflector, and space apparatus. The plasma injector creates and accelerates the ultra-cold low density plasma.

The Installation works the following way:  the injector-accelerator forms and injects the cold neutral plasma beam with high speed in spaceship direction. When the beam reaches the ship, the electrostatic collector of spaceship collects and separates the beam ions from large area and passes them through the engine-electric generator or reflects them by electrostatic mirror. If we want to receive the thrust in the near beam direction ($\pm 90^\circ$) and electric energy, the engine works as thruster (accelerator of spaceship and braker of beam) in beam direction and electric generator. If we want to get thrust in opposed beam direction, the space engine must accelerate the beam ions and spend energy. If we want to have maximum thrust in beam direction, the engine works as full electrostatic mirror and produces double thrust in the beam direction (full reflection of beam back to injector). The engine does not spend energy for full reflection.

The thrust is controlled by the electric voltage between engine nets [19], the thrust direction is controlled by the engine nets angle to beam direction. Note, the trust can brake the ship (decrease the tangential ship speed) and far ship (located out of Earth orbit) can return to the Earth by Sun gravity. Note also, the Earth atmosphere absorbs and scatters the plasma beam and the beam injector must be located on Earth space mast or tower (up $40 \div 60$ km, see [20, 21]) or the Moon. Only high energy beam can break through atmosphere with small divergence. The advantage: the injector has a reflector and when the ship locates not far from the injector the beam will be reflected a lot of times and thrust increases in thousand times at start (fig. 2) (see same situation in [22]).

The proposed engine may be also used as AB-ramjet engine [19], utilizing the Solar wind or interstellar particles.

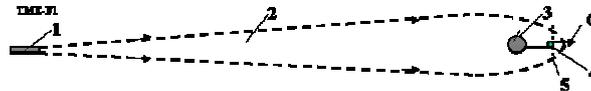

**Fig.1**. Long distance space transfer of electric energy, matter, and momentum (thrust). Notation are: 1 - injector-accelerator of neutral ultra-cold plasma (ions and electrons), 2 - plasma beam, 3 - space ship or planetary team, 4 - electrostatic ions collector (or magnetic collector), 5 - braking electric nets (electrostatic electro-generator-thruster-reflector), 6 - thrust.

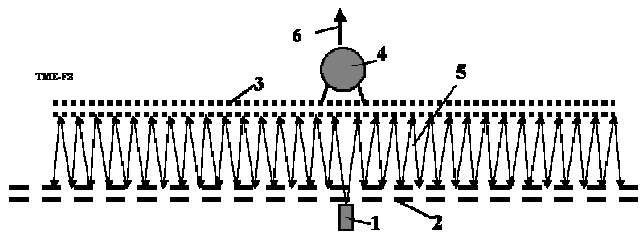

**Fig.2.** Multi-reflection start of the spaceship having proposed engine. Notation are: 1 - injector-accelerator of cold ions or plasma, 2 , 3 -  electrostatic reflectors, 4 - space ship, 5 - plasma beam, 6 - the thrust.

The electrostatic collector and electrostatic generator-thruster-reflector proposed and described in [19]. The main parts are presented below.

A **Primary Ramjet** propulsion engine is shown in fig.3. Such an engine can work in charged environment.  For example, the surrounding region of space medium contains positive charge particles (protons, ions). The engine has two plates 1, 2, and a source of electric voltage and energy (storage) 3. The plates are made from a thin dielectric film covered by a conducting layer. The plates may be a net. The source can create an electric voltage $U$ and electric field (electric intensity $E$) between the plates. One also can collect the electric energy from plate as an accumulator.

The engine works in the following way. Apparatus are moving (in left direction) with velocity $V$ (or particles 4 are moving in right direction). If voltage $U$ is applied to the plates, it is well-known that main electric field is only between plates. If the particles are charged positive (protons, positive ions) and the first and second plate are charged positive and negative, respectively, then the particles are accelerated between the plates and achieve the additional velocity $v > 0$. The total velocity will be $V+v$ behind the engine (fig.3a). This means that the apparatus will have thrust $T > 0$ and spend electric energy $W < 0$ (bias, displacement current). If the voltage $U = 0$, then $v = 0$, $T = 0$, and $W = 0$ (fig.3b).

If the first and second plates are charged negative and positive, respectively, the voltage changes sign. Assume the velocity $v$ is satisfying $-V < v < 0$. Thus the particles will be braked and the engine (apparatus) will have drag and will also be braked. The engine transfers braked vehicle energy into electric (bias, displacement) current. That energy can be collected and used. Note that velocity $v$ cannot equal $-V$. If $v$ were equal to $-V$, that would mean that the apparatus collected positive particles, accumulated a big positive charge and then repelled the positive charged particles.

If the voltage is high enough, the brake is the highest (fig.3d). Maximum braking is achieved when $v = -2V$ ($T < 0$, $W = 0$). Note, the $v$ cannot be more then $-2V$, because it is full reflected speed.

**AB-Ramjet engine**. The suggested Ramjet is different from the primary ramjet. The suggested ramjet has specific electrostatic collector 5 (fig. 4a,c,d,e,f,g). Other authors have outline the idea of space matter collection, but they did not describe the principal design of collector. Really, for charging of collector we must move away from apparatus the charges. The charged collector attracts the same amount of the charged particles (charged protons, ions, electrons) from space medium. They discharged collector, work will be idle. That cannot be useful.

The electrostatic collector cannot adsorb matter (as offered some inventors) because it can adsorb ONLY opposed charges particles, which will be discharged the initial charge of collector. Physic law of conservation of charges does not allow to change charges of particles.

The suggested collector and ramjet engine have a special design (thin film, net, special form of charge collector, particle accelerator). The collector/engine passes the charged particles ACROSS (through) the installation and changes their energy (speed), deflecting and focusing them. That is why we refer to this engine as the ***AB-Ramjet engine***. It can create thrust or drag, extract energy from the kinetic energy of particles or convert the apparatus' kinetic energy into electric energy, and deflect and focus the particle beam. The collector creates a local environment in space because it deletes (repeals) the same charged particles (electrons) from apparatus and allows the Ramjet to work when the apparatus speed is close to zero. The author developed the theory of the electrostatic collector and published it in [23]. The conventional electric engine cannot work in usual plasma without the main part of the AB-engine - the special pervious electrostatic collector.

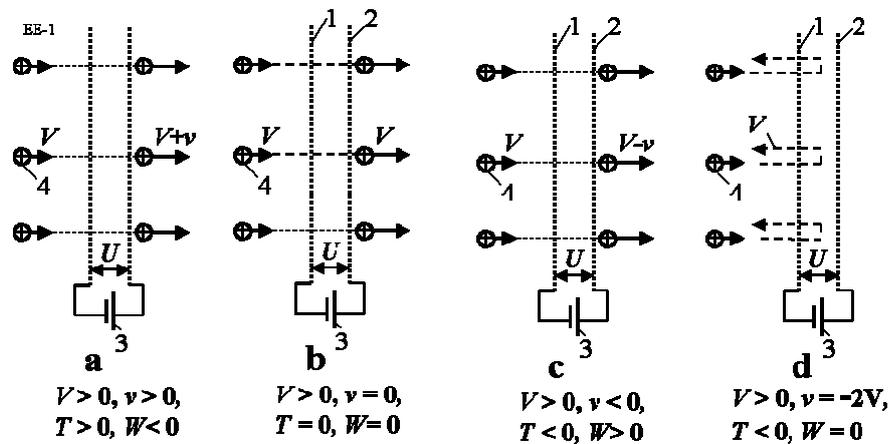

**Fig. 3**. Explanation of primary Space Ramjet propulsion (engine) and electric generator (in braking),
a) Work in regime *thrust*; b) Idle; c) Work in regime *brake*. d) Work in regime *strong brake (full reflection)*.
Notation: 1, 2 - plate (film, thin net) of engine; 3 - source of electric energy (voltage $U$); 4 - charged particles (protons, ions); $V$ - speed of apparatus or particles before engine (solar wind); $v$ - additional speed of particles into engine plates; $T$ - thrust of engine; $W$ - energy (if $W < 0$ we spend energy) .

The plates of the suggested engine are different from the primary engine. They have concentric partitions which create additional radial electric fields (electric intensity) (fig. 4b). They straighten, deflect and focus the particle beams and improve the efficiency coefficient of the engine.

The central charge can have a different form (core) and design (fig.4 c,d,e,f,g,h). It may be:
1) a sphere (fig. 4c) having a thin cover of plastic film and a very thin (some nanometers) conducting layer (aluminum), with the concentric spheres inserted one into the other (fig.4d),
2) a net formed from thin wires (fig. 4e);
3) a cylinder (without butt-end)(fig.4f); or
4) a plate (fig.4g).

The design is chosen to produce minimum energy loss (maximum particle transparency - see section "Theory"). The safety (from discharging, emission of electrons) electric intensity in a vacuum is $10^8$ V/m for an outer conducting layer and negative charge. The electric intensity is more for an inside conducting layer and thousands of times more for positive charge.

The engine plates are attracted one to the other (see theoretical section). They can have various designs (fig.5a - 5d). In the rotating film or net design (fig.5a), the centrifugal force prevents contact between the plates. In the inflatable design (fig. 5b), the low pressure gas prevents plate contact. A third design has (inflatable) rods supporting the film or net (fig. 5c). The fourth design is an inflatable toroid which supports the distance between plates or nets (fig. 5d).

Note, the AB-ramjet engine can work using the neutral plasma. The ions will be accelerated or braked, the electrons will be conversely braked or accelerated. But the mass of the electrons is less then the mass of ions in thousands times and AB-engine will produce same thrust or drag.

**Plasma accelerator.** The simplest linear plasma accelerator (principle scheme of linear particle accelerator) for plasma beam is presented in fig. 6. The design is a long tube (up 10 m) which creates a strong electric field along the tube axis (100 MV/m and more) . The accelerator consists of the tube with electrical isolated cylindrical electrodes, ion source, and voltage multiplier. The accelerator increases speed of ions, but in end of tube into ion beam the electrons are injected. This plasma accelerator can accelerate charged particles up 1000 MeV. Electrostatic lens and special conditions allow the creation of a focusing and self-focusing beam which can transfer the charge and energy long distances into space. The engine can be charged from a satellite, a spaceship, the Moon, or a top atmosphere station (space tower [21 - 22]). The beam may also be used as a particle beam weapon.

Approximately ten years ago, the conventional linear pipe accelerated protons up to 40 MeV with a beam divergence of $10^{-3}$ radian. However, acceleration of the multi-charged heavy ions may result in significantly more energy.

At present, the energy gradients as steep as 200 GeV/m have been achieved over millimeter-scale distances using laser pulsers. Gradients approaching 1 GeV/m are being produced on the multi-centimeter-scale with electron-beam systems, in contrast to a limit of about 0.1 GeV/m for radio-frequency acceleration alone. Existing electron accelerators such as SLAC <http://en.wikipedia.org/wiki/SLAC> could use electron-beam afterburners to increase the intensity of their particle beams. Electron systems in general can provide tightly collimated, reliable beams while laser systems may offer more power and compactness.

The cool plazma beam carries three types of energy: kinetic energy of particles, ionization, and dissosiation energy of ions and moleculs. That carry also particle mass and momentum. The AB-Ramjet engine (discribed over) can utilise only kinetic energy of plasma particles and momentum. The particles are braked and produce an electric current and thrust or reflected and produce only thrust in the beam direction. If we want to collect a plasma matter and to utilize also the ionization energy of plasma (or space invironment) ions and dissociation energy of  plasma molecules we must use the modified AB-Ramjet engine discribed below (fig.7).

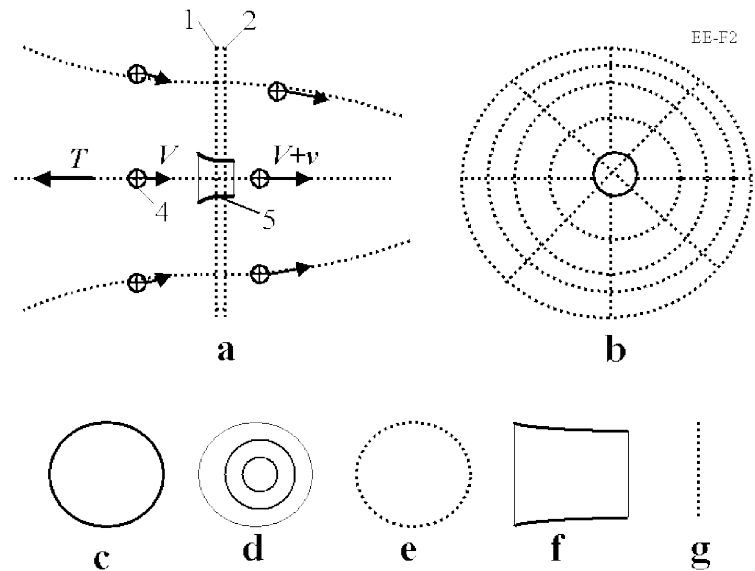

**Fig.4.** Space AB-Ramjet engine with electrostatic collector (core). a) Side view; b) Front view; c) Spherical electrostatic collector (ball); d) Concentric collector; e) cellular (net) collector; f) cylindrical collector without cover butt-ends; g) plate collector (film or net).

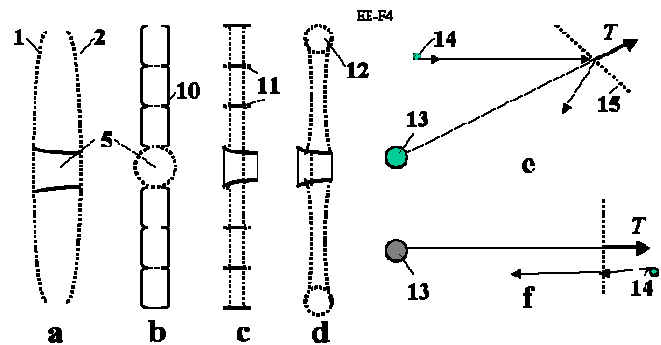

**Fig. 5.** Possible design of the main part of ramjet engine. a) Rotating engine; b) Inflatable engine (filled by gas); c) Rod engine; d) Toroidal shell engine, e) AB-Ramjet engine in brake regime, f) AB-Ramjet engine in thrust regime. Notation: 10 - film shells (fibers) for support thin film and creating a radial electric field; 11 - Rods for a support the film or net; 12 - inflatable toroid for support engine plates; 13 - space apparatus; 14 - particles; 15 - AB-Ramjet.

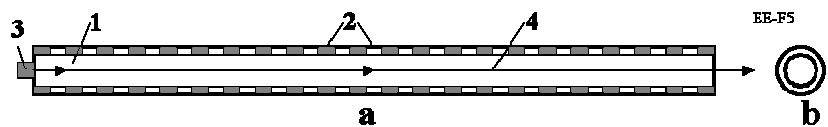

**Fig. 6.** Plasma accelerator for transfer charges (energy) in long distance space ship or for charging AB-Ramjet engine. a) Side view, b) Front view. Notations: 1 - gun tube, 2 - opposed charged electrodes, 3 - source of charged particles (ions, electrons), 4 - particles beam.

The modified AB-engine has magnetic collector (option), three nets (two last nets may be films), and issue voltage (that also may be an electric load). The voltage, $U$, must be enough for full braking of charged particles. The first two nets brake the electrons and precipitate (collect) the electrons on the film 2 (fig.7). The last couple of film (2, 3 in fig.7) brakes and collects the ions. The first couple of nets accelerate the ions that is way the voltage between them must be double.

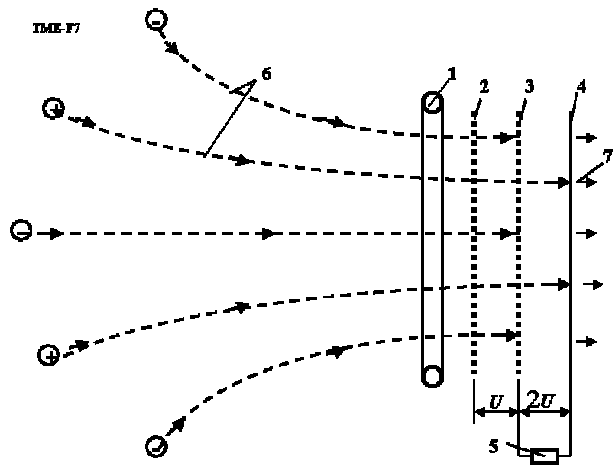

**Fig.7.** AB-engine which collected matter of plasma beam, kinetic energy of particles, energy ionization and dissociation. Notations: 1 - magnetic collector; 2 - 4 - plates (films, nets) of engine; 5 - electric load; 6 - particles of plasma; 7 - radiation. *U* - voltage between plates (nets).

The collected ions and electrons have the ionized and dissociation energy. This energy is significantly (up 20 - 150 times) more powerful then chemical energy of rocket fuel (see Table 1) but significantly less then kinetic energy of particles (ions) equal *U* (in eV) (*U* may be millions volts). But that may be used by ship. The ionization energy conventionally pick out in photons (light, radiation) which easy are converted in a heat (in closed vessel), the dissociation energy conventionally pick out in heat.

The light energy may be used in the photon engine as thrust (fig. 8a) or in a new power laser (fig.8b). The heat energy may be utilized conventional way (fig. 8c). The offered new power laser (fig. 8b) works the following way. The ultra-cool rare plasma with short period of life time located into cylinder. If we press it (decrease density of plasma) the electrons and ions will connect and produce photons of very closed energy (laser beam). If we compress very quickly by explosion the power of beam will be high. The power is only limited amount of plasma energy.

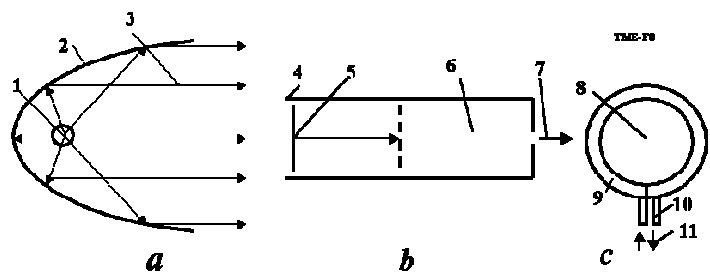

**Fig.8**. Conversion of ionization energy into radiation and heat. *a* - photon engine; *b* - power laser (light beamer); *c* - heater. Notations: 1 - recombination reactor; 2 - mirror; 3 - radiation (light) beam; 5 - piston; 6 - volume filled by cold rare plasma; 7 - beam; 8 - plasma; 9 - heat exchanger; 10 - enter and exit of hear carrier; 11 - heat carrier.

After recombination ions and electrons we receive the conventional matter. This matter may be used as nuclear fuel (in thermonuclear reactor), medicine, food, drink, oxidizer for breathing, etc.

## Transfer Theory of the high speed neutral ultra-cold plasma and particles

Below are the main equations and computations of neutral ultra-cold plasma beam having velocity up to relativistic speed. These equations received from conventional mechanics and relativistic theory.

Note a ratio $\beta$

$$\beta = \frac{V}{c}, \quad \beta_s = \frac{V_s}{c},\qquad(1)$$

where $V$ is plasma beam speed, m/s; $c = 3 \times 10^8$ is light speed, m/s; $V_s$ is projection of a ship speed in beam direction.

1. **Relative relativistic time**, $\bar{t}$, for observer moving together with beam is

$$\bar{t}' = \frac{t'}{t} = \sqrt{1 - \beta^2},\qquad(2)$$

where $t'$ is time for observer moving together with beam (system coordinate connected with beam)[s], $t$ is time for Earth's observer [s]. Computation of Eq. (2) is presented in Fig. 9. The beam time decreases for relativistic speed. That means the beam divergence is also decreased and beam energy may be passed for long distance.

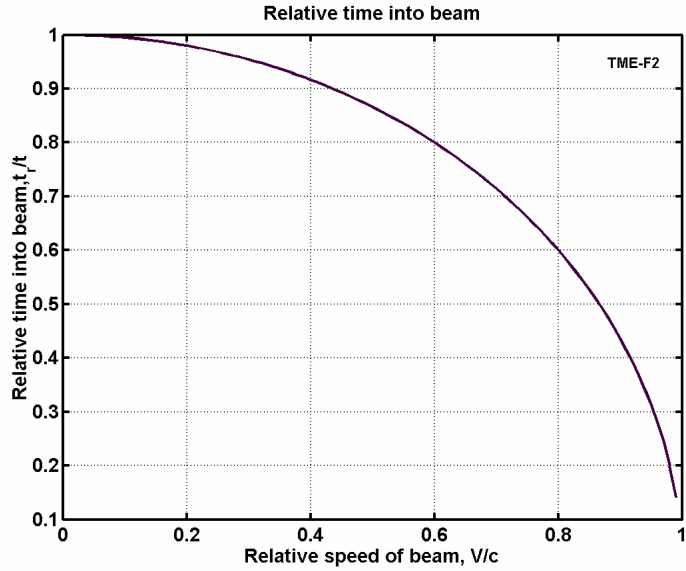

**Fig.9.** Beam relative time versus beam relative speed for high relativistic beam speed.

2. **The power spent for acceleration plasma beam** in Earth for efficiency = 1 (kinetic power of particle beam) is

$$P_B = \frac{M_0 c^2}{2}\frac{\beta^2}{\sqrt{1 - \beta^2}}, \quad \text{or} \quad \text{for} \quad \beta \ll 1 \quad P_B = \frac{M_0 V^2}{2}, \quad [\text{W/s}]\qquad(3)$$

where $M_0$ is mass flow of beam, kg/s in Earth system of coordinate.

The computations of Eq. (3) for the intervals $(0 \div 0.1)c$ and $(0 \div 0.95)c$ are presented in Figs. 10, 10a. The relativistic speed needs very high power in any method because the relativistic beam requires this energy.

3. **The power $P_i$ of dissociation and single ionization** of one nucleon is

$$P_i = 1.6 \times 10^{-19}\frac{M_0}{m_p n}e_i \quad [J] \quad \text{or} \quad P_i = \frac{M_0}{m_p n}e_i \quad [\text{eV}],\qquad(4)$$

where $m_p = 1.67 \times 10^{-27}$ kg is mass of proton, $n$ is number of nucleon in nucleus, $e_i$ is energy of dissociation, ionization, or molecular breakup respectively. The energy of the first ionization (ion lost one electron) approximately equals from 2 to 14 eV. Magnitudes of this energy for some molecules and ions are in Table No.1.

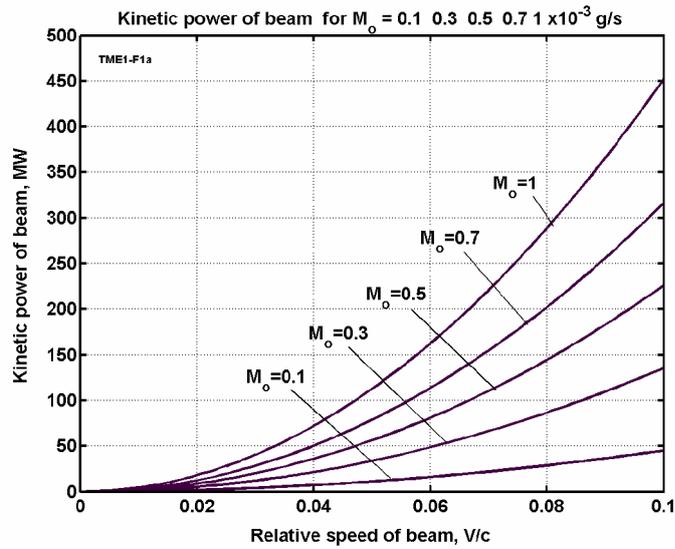

**Fig. 10**. Power for the beam acceleration via beam flow mass and relative beam speed for interval
$(0 \div 0.1)c$

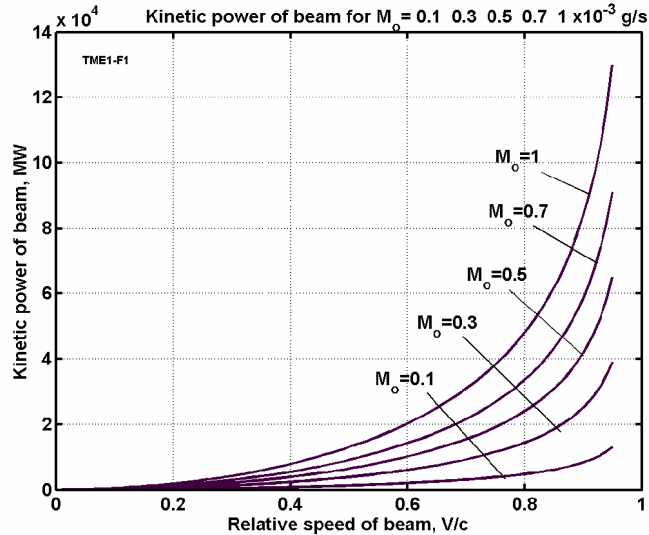

**Fig. 10a**. Power for the beam acceleration via beam flow mass and relative beam speed for interval
$(0 \div 0.95)c$

**Table 1.** Energy ionization, dissociation, and molecular breakup of some molecules and ions in $eV$

===================================================

| | | |
|---|---|---|
| Molecular breakup | $H_2O \rightarrow H_2 + O$  2 eV, | $CO_2 \rightarrow C + O_2$   0.093 eV |

-------------------------------------------------

| | | |
|---|---|---|
| Dissociation | $H_2 \rightarrow H + H$  4.48 eV | $O_2 \rightarrow O + O$   5.1 eV |

-------------------------------------------------

| | | |
|---|---|---|
| Ionization | $H \rightarrow H^+$  13.6 eV | $H_2 \rightarrow H_2^+$  2.65 eV  $O_2 \rightarrow O_2^+$  6.7 eV |

===================================================

If speed is relativistic, this energy is small in comparison with kinetic energy of beam. For interplanetary speed ($V_S$ = 8 -15 km/s) the energy of ionization reaches 15 - 50% from kinetic energy of beam. That decreases the coefficient of efficiency launch installation. If we used the heavy ions or a charged matter, the ionization energy decreases but voltage increases. For interplanetary vehicles it is not important because required voltage for low speed are small ($U \approx 5 \div 20$ V).
Fig. 11 shows the required energy for different case

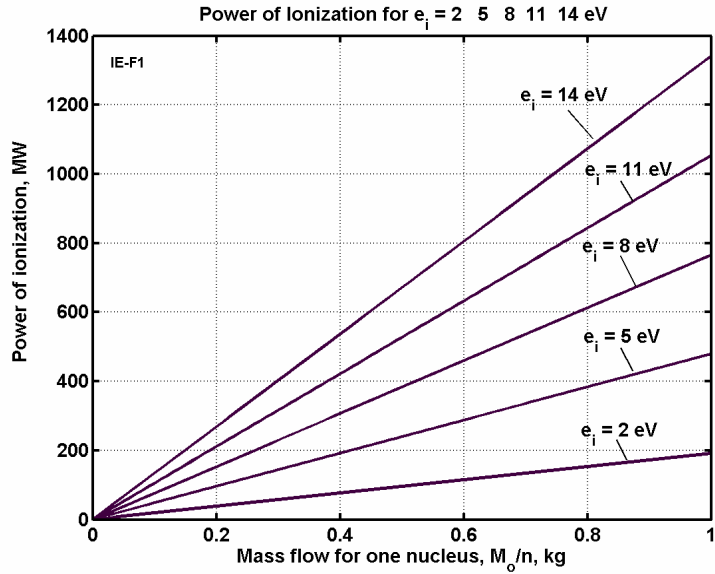

**Fig. 11.** Power of ionization versus mass flow and ionization potencial (Eq. (4)).

**4**. **The maximal thrust (drag)** from the full reflected one charged plasma beam, for Earth's observer and relativistic speed and non-relativistic speed may be estimated by following equations:

$$T_{max} = 2M(V \mp V_S) \approx \frac{2M_0 c(\beta \mp \beta_S)}{\sqrt{1 - \beta^2}}, \quad \text{or} \quad \text{for} \quad \beta_S \ll 1 \quad \text{the thrust is} \quad T_{max} = \frac{2M_0 c\beta}{\sqrt{1 - \beta^2}},$$

$$\text{for} \quad \beta \ll 1, \quad \beta_S \ll 1, \quad \text{the thrust is} \quad T_{max} = 2M_0(V - V_S) \quad (5)$$

Here $M$ is calculated mass of a moving relativistic particle flow, kg/s; $M_0$ is mass of the particle flow measured by Earth's observer, kg/s.

*Note*: If the space ship move along the beam in same direction, the thrust is decreased (sign is "-"); if that moves in opposed direction, the drag is increased (sing is "+"). This drag (thrust) is not requested the ship propulsion energy.

Result of computation for intervals $(0 \div 0.1)c$, $(0 \div 0.95)c$, $V_S = 0$ are presented in Fig. 12, 12a.

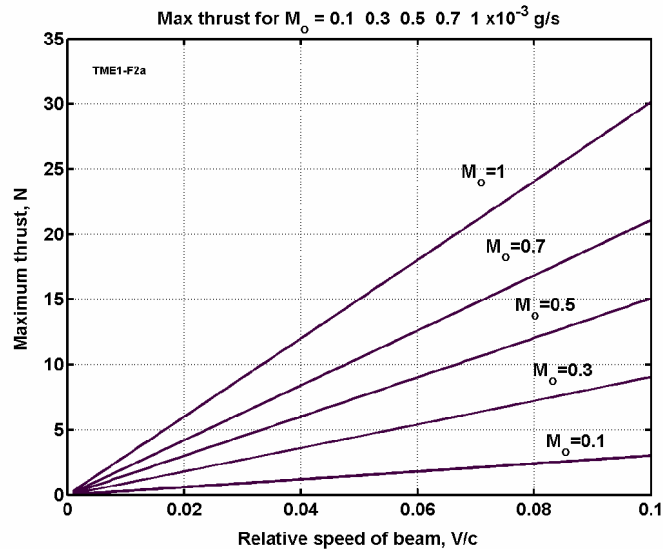

**Fig. 12**. Maximum thrust (drag) is produced by beam in space ship for $V_S = 0$ and the interval

$(0 \div 0.1)c$

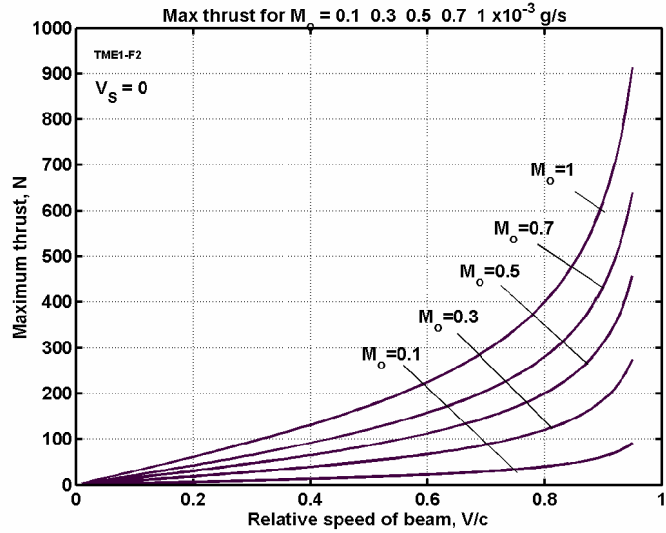

**Fig. 12a**. Maximum thrust (drag) is produced by beam in space ship for $V_S = 0$ and interval
$(0 \div 0.95)c$

**5**. **The divergence of beam** is a very important magnitude. If divergence is small, we can pass our energy in long distance $S$:

$$D = \frac{ut'}{Vt} = \frac{u}{c}\frac{\sqrt{1-\beta^2}}{\beta}S, \quad \overline{D} = \frac{D}{S}, \quad S = c\beta t \,, \tag{6}$$

where $u$ is maximal radial speed, m/s; $D$ is maximal radial distance (radius of plasma beam), m; $\overline{D}$ is relative divergence (angle of divergence, $\theta = 2\overline{D}$ radians); $t$ is time of beam moving, s.

The computation of Eq. (6) is shown in Fig. 13. We need in small $u$ (ultra-cold plasma) for decreasing of divergence as small as possible ($u = 0.01 - 1$ m/s). In this case we can transfer energy in the large distance and accelerate a ship for relativistic speed. The plasma is mixture of ions and electrons. If it is low-density, it can exist a long time. The cold plasma can be emitted from solid electrodes.

*Note:* Equations (2),(6) shows when $V \to c$, then $t' \to 0$ and deviation $D \to 0$. That means the deviation can be small as we want but we need a big power for it.

The corresponding temperature is

$$T_c = \frac{mu^2}{ik} \,, \tag{7}$$

where $m$ is mass of molecule (ion) [kg]; $m = m_p n$, here $m_p = 1.67 \times 10^{-27}$ is mass of proton, $n$ is number of nucleons into nucleus; $i = 3$ for single ion (for example $O^+$), $i = 5$ for double molecule (for example $O_2^+$), $i = 6$ for multi-molecular ions, $k = 1.39 \times 10^{-23}$ is Boltzmann constant.

For $u = 0.1 \div 1$ m/s the temperature is about $10^{-3}$ °K, the relative divergence is $10^{-9}$.

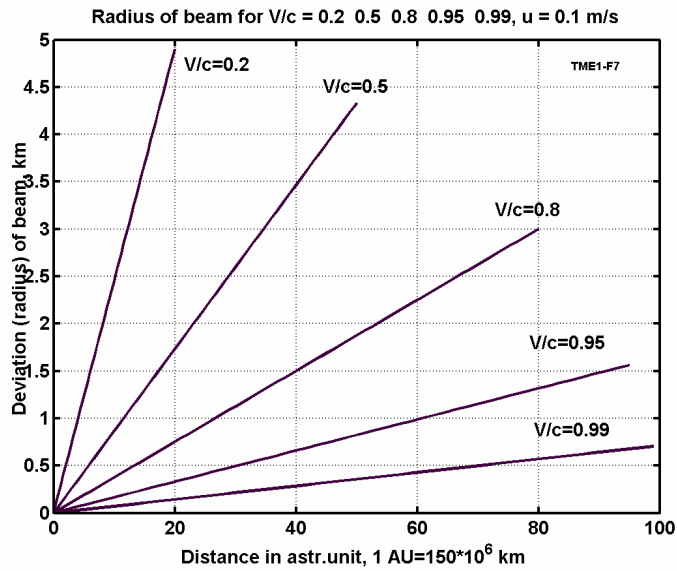

**Fig. 13**. Radius of beam divergence via distance and ratio *V/c* .

5. **Accelerate voltage** is

$$U = \frac{mV^2}{2q} = \left(\frac{m_p}{q}\right)\frac{nc^2}{2}\frac{\beta^2}{\sqrt{1-\beta^2}} \; ,$$ (8)

where $q = 1.6 \times 10^{-19}$ C is electron (ion) charge. The computations are presented in Fig. 14, 14a. The need voltage may be reduced in *Z* times if the ion has *Z* charges (delete *Z* electrons from ion).

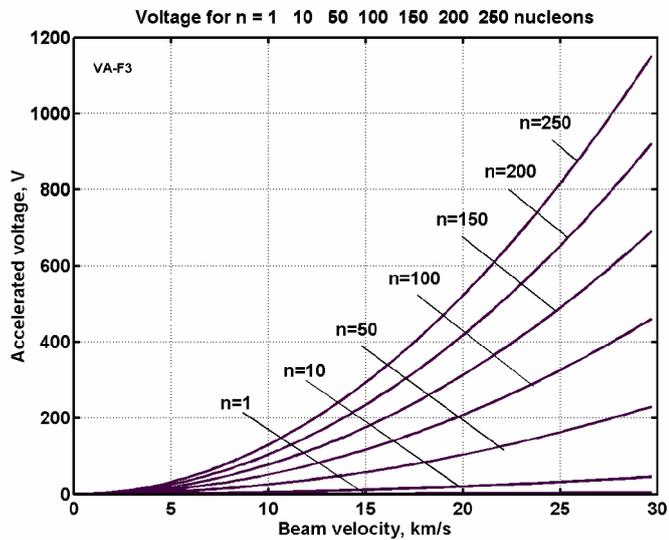

**Fig. 14**. Accelerated voltage versus the coventinal beam speed and number of nucleons.

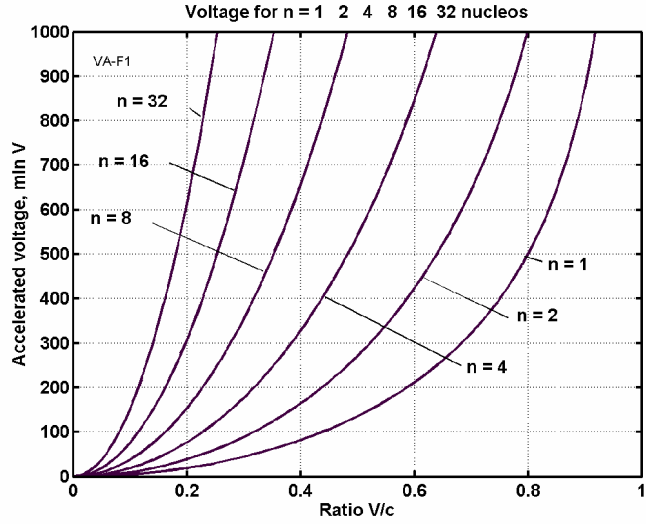

**Fig. 14a**. Accelerated voltage versus a relativistic speed ratio $V/c$ and number of nucleons.

**6. The speed $V_s$ and distance of space ship $S$** can be computed by conventional method (Earth's observer):

$$V_S = at, \quad S = \frac{at^2}{2}, \quad S = \frac{V_S^2}{2a}, \quad a = \frac{T}{M_S},$$

$a$ is ship acceleration, m/s$^2$. $M_s$ is ship mass, kg, $V_S$ is ship speed measured by Earth's observer, m/s .

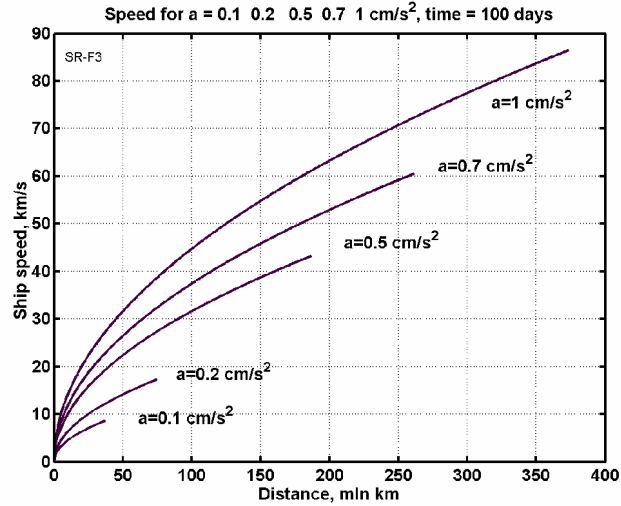

**Fig.15**. Ship speed for 100-days flight versus distance and ship acceleration.

**7. Relative beam speed** for a ship observer is

$$\beta_{BS} = \frac{\beta \pm \beta_S}{1 + \beta \beta_S}, \tag{9}$$

where $\beta$, $\beta_S$ is relative speed of beam and space ship respectively measured by Earth's observer. The sign " - " is used for same direction of speeds.

**8. Loss energy of the beam** in the Earth atmosphere may be estimated by the following way:

$$\tau = \frac{100 H_0 \rho_0 \overline{\rho}(h) \overline{p}(h)}{R_i(U)}, \quad R_t = \frac{m}{m_p} R_i\left(\frac{m_p}{m} U\right), \tag{10}$$

where $H_0 = P_a/\rho = 10^4/1.225 = 8163$ m is thickness (height) of Earth atmosphere having constant density $\rho = 1.225$ kg/m$^3$, $P_a = 10^4$ kg/m$^2$ is the atmospheric pressure; $\overline{\rho}(h)$ is relative atmosphere density; $\overline{p}(h)$ is relative atmosphere pressure; $R_t$ is particle track in atmosphere [cm]; $m$ is mass of particle, kg; $h$ is altitude, m; $U$ is beam energy, MeV; $\rho_0 = 0{,}001225$ g/cm$^3$ is atmosphere density; 100 is transfer coefficient meter into cm. Magnitudes $R_t$, $\overline{\rho}(h)$, $\overline{p}(h)$ for proton are given below in Table 2, 3.

**Table 2.** Value $R_t$ [g/cm$^2$] versus energy of proton in MeV, [32], p. 953.

| $U$ MeV | 0.1 | | 1 | | 10 | | 50 | 100 | 200 | 300 |
|---|---|---|---|---|---|---|---|---|---|---|
| $R_t$ g/cm$^2$ | $1\times10^{-4}$ | | $1.09\times10^{-2}$ | | $0{,}99\times10^{-1}$ | | 2.56 | 8.835 | 29.64 | 58.08 |
| $U$ | 400 | 500 | 600 | 700 | 800 | 1000 | 2000 | 3000 | 5000 | 7000 | 10,000 |
| $R_t$ | 93.73 | 133.3 | 176 | 222 | 270 | 370 | 910 | 1363 | 2543 | 3583 | 5081 |

**Table 3.** Standard Earth atmosphere, [33], p. 261.

| $h$ km | 0 | 5 | 10 | 20 | 40 | 60 | 100 |
|---|---|---|---|---|---|---|---|
| $\overline{\rho}(h)$ | 1 | 0.661 | 0.338 | 0.072 | $3.27\times10^{-3}$ | $2.71\times10^{-4}$ | $4.41\times10^{-7}$ |
| $\overline{p}(h)$ | 1 | 0.533 | 0.261 | 0.054 | $2.92\times10^{-3}$ | $8.35\times10^{-4}$ | $3.20\times10^{-7}$ |

Results of computation Eq. (10) are presented in Fig. 16, 16a, 16b.

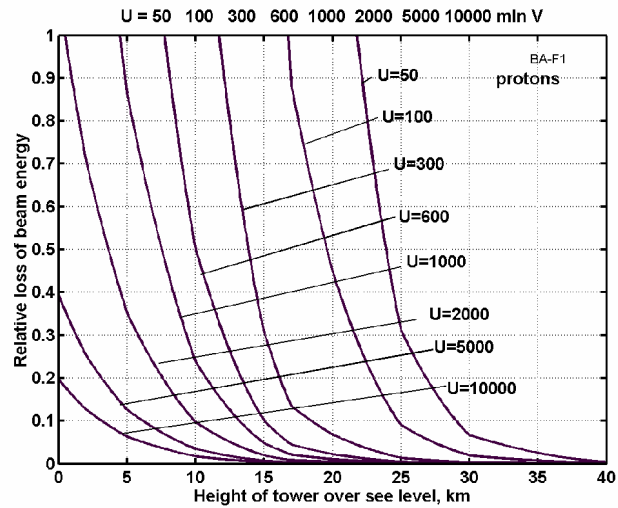

**Fig.16.** Relative energy loss of the proton particle beam via a tower altitude in Earth atmosphere. Accelerate voltage $U$ are in millions volts.

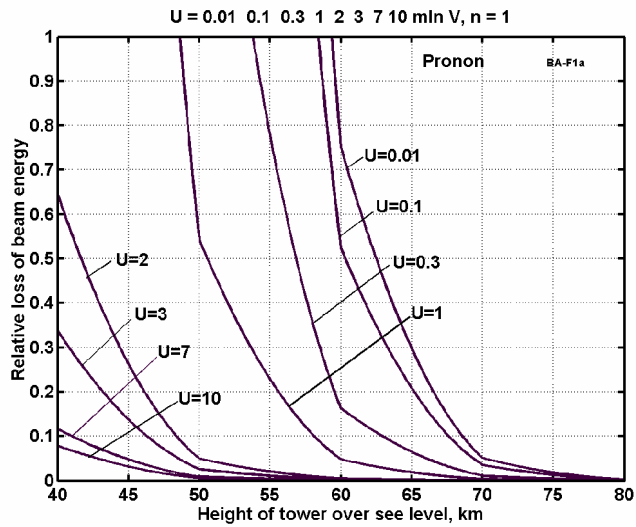

**Fig.16a.** Relative energy loss of the proton particle beam via a tower altitude in Earth atmosphere. Accelerate voltage $U$ are in millions volts. Angles in curve are result of the linearization data of Table 2, 3.

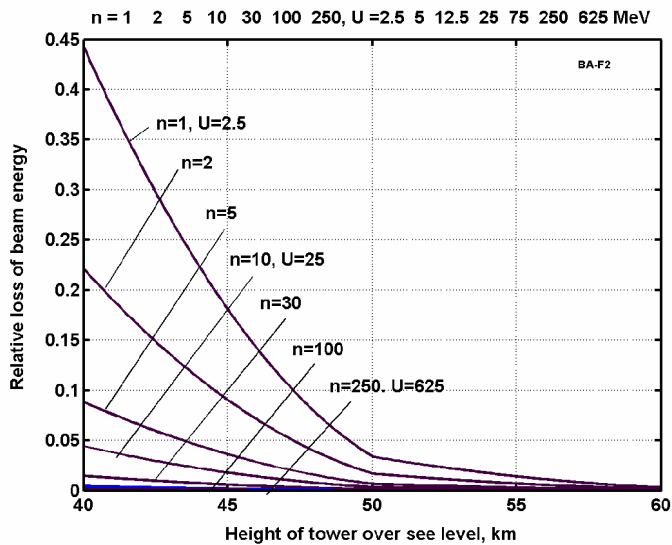

**Fig.16b.** Relative energy loss of the particle beam via the tower altitude in Earth atmosphere and number $n$ of nucleons in nucleus. Accelerate voltage $U$ in millions volts.

Evidently, only high energy particle beam break up the Earth atmosphere. There is no problem if the particle beam starts from a space tower [20], [25] of 40 ÷ 80 km altitude or from the Moon.

Last formula in (10) allows recalculation by the particle track for any atom. For example, we want to calculate the particle track for oxidizer particle having $m = 16m_p$ and energy 8,000 MeV. We take the $R_t$ from Table 2 for $U = 8,000/16 = 500$ MeV and multiple by 16. Result $R_t = 133 \times 16 = 2128$. The particle track $T_r = R_t/\rho_o = 2128/0.001225 = 1737142$ cm $= 17.4$ km in the air having density 1.225 kg/m$^3$. That is enough to break the Earth's atmosphere of the constant density 8.163 km, but the loss of energy will be $\tau = 8.163/17.4 = 0.47$ (47%). The divergence may also be increased by atmosphere. Loss and divergence may be improved is the beam station is located on a mountain or special tower having the height about 40 ÷ 60 km .

**9. Multi-reflex launch and landing** (Fig.2). In a starting or braking period the thrust (braking) can be increasing if we use the multi-reflect method developed in [26]. Multi-reflect in launching does not increased the installation power (thrust is increased by increasing of efficiency), multi-reflex in braking converts the apparatus kinetic energy into the electric energy which can be utilized by apparatus or operated station. The theory of multi-reflection is described below (see also [26]).

*Change in beam power*. The beam power will be reduced if one (or both) reflector is moved, because the beam speed changes. The total relative loss, $q$, of the beam energy in one double cycle (when the beam is moved to the reflector and back) is

$$q = (1–2\gamma)(1–2\xi)(1\pm2v)\varsigma , \quad q > 0 , \qquad (11)$$

where $v$ is the loss (useful work) through relative mirror (lens) movement, $v = V_S/V$, $V_S$ is the relative speed of the electrostatic mirrors (space apparatus)[m/s], $V =$ is the speed of the beam (in system of coordinates connected with an power operating station). We take the "+" when the distance is reduces (braking) and take "– " when the distance is increased (as in launching, a useful work for beam), $\gamma$ is coefficient reflectivity of electrostatic mirror (the loss of beam energy through the electrostatic reflector); $\xi$ is the loss (attenuation) in the medium (air) (see point 8). If no atmosphere, $\xi = 0$; $\varsigma$ is the loss through beam divergence ($\varsigma = 1$ if $D < D_r$, where $D_r$ is diameter of the electrostatic mirror). For a wire net electrostatic reflector $\gamma \approx 2d_w/l$ where $d_w$ is diameter of wire, $l$ is size of mesh. For example, for the net having a mesh $0.1\times0.1$ m, the $l = 0.1$ m and a wire $d_w = 0.0001$ m, the $\gamma = 0.002$ .

*Multi-reflex light pressure*. The beam pressure, $T$, of two opposed high reflectors after a series of reflections, $N$, to one another is

$$T(V_S) = T_0 + \frac{2P_B}{V}\sum_{j=1}^{N} q^j(V_S), \quad N(V_S) = \sqrt{\frac{kD_rV}{uS}}. \quad P_B = \frac{M_0V^2}{2}. \quad T_0 = 2M_0(V - V_S) \geq 0 , \qquad (12)$$

where $S$ is distance between electrostatic reflectors of the station and ship [m], $k = 1 \div 1.5$ is correction coefficient for the case when $D > D_r$. For primary estimation $k = 1$.

If $V_S$ is small and $V$ is high, the multi-reflex $T$ may be large. For example, if $V_S =$10 m/s, $V =$ 30 km/s, $S$ is small, the number of reflection may reach $n \approx 30000/10/2 = 1500$ times more then regular thrust. That is well for ship trip starting and braking.

*Limitation of reflection number*. If the reflector is moved away, the maximum number of reflections, $N$, is limited by $q > 0$, $V_S < 0.5 V$ (see Eq. (11)). At ship launch or braking the maximum thrust is limited by a safety acceleration or deceleration.

*Coefficient of efficiency*. The propulsion efficiency coefficient, $\eta$, (without loss for ionization) may be computed using the equation

$$\eta = TV_S / P_B , \qquad (13)$$

For full reflection Eq. (13) has the form:

$$\eta = \frac{4(V - V_S)V_S}{V^2}, \quad \eta_{max} = 1 \quad \text{for} \quad V = 2V_S . \qquad (14)$$

Computation of launch and landing trajectories computed by the usual method of integration

$$a_i = T / M_S , \quad t_{i+1} = t_i + \Delta t, \quad V_{S,i+1} = V_{S,i} + a_i\Delta t, \quad S_{i+1} = S_i + V_{S,i}\Delta t \qquad (15)$$

The results for spaceship having the weight 3 tons and the final speed 12 km/s are presented in Fig. 17 - 19.

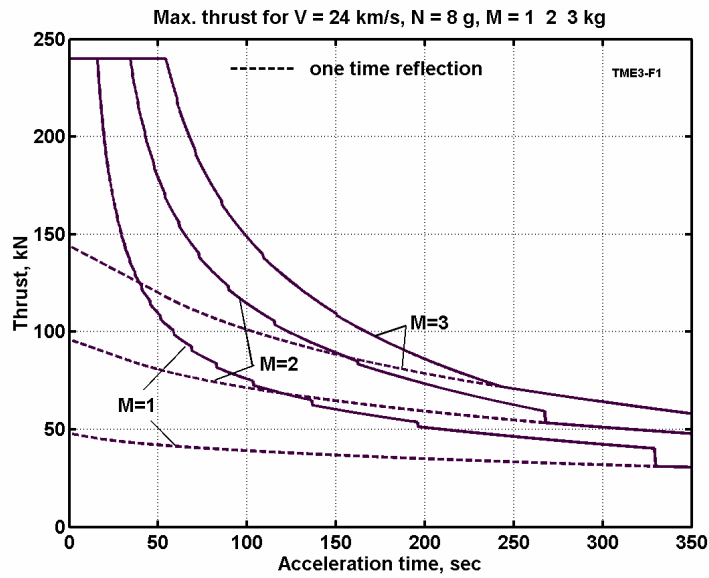

**Fig. 17.** Thrust of the multi-reflection beam and the one time reflection beam versus the flight time and beam flow. Data: $\gamma = 0.002$, $\xi = 0$, $\zeta = 1$, $D_r = 30$ m, $u = 0.1$ m, $V = 25000$ m/s, $M_s = 3000$ kg, acceleration limit $N_g < 8$g, $M_0 = M = 1, 2, 3$ kg/s.

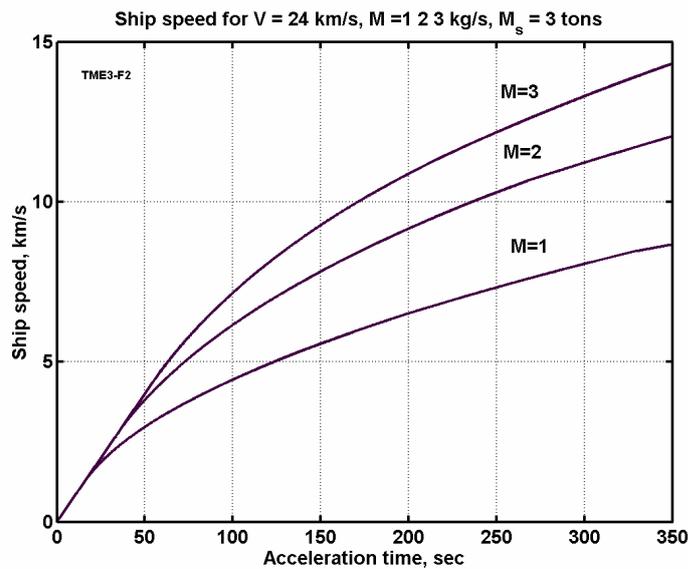

**Fig. 18.** Ship speed versus the flight-time and the beam flow. Data: $\gamma = 0.002$, $\xi = 0$, $\zeta = 1$, $D_r = 30$ m, $u = 0.1$ m, $V = 25000$ m/s, $M_s = 3000$ kg, acceleration limit $N_g < 8$g, $M_0 = M = 1, 2, 3$ kg/s.

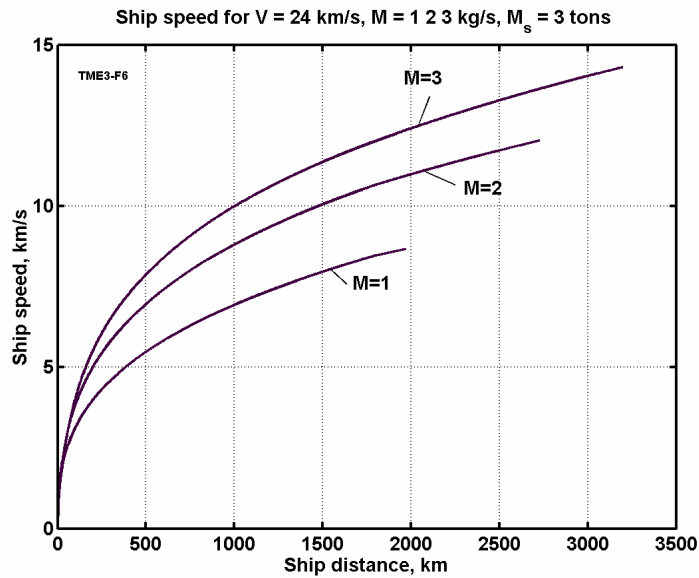

**Fig. 19.** Ship speed versus the distance and the beam flow. Data: flight time $t$ = 350 sec., $\gamma$ = 0.002, $\xi$ = 0, $\zeta$ = 1, $D_r$ =30 m, $u$ = 0.1 m, $V$ = 25,000 m/s, $M_s$ = 3000 kg, acceleration limit $N_g$< 8g, $M_0$= $M$ = 1, 2, 3 kg/s.

**9.** For **non-relativistic flight** all equations are simplify.

$$T_{\max} = 2M_0\left(V - V_S\right), \quad P_B = \frac{M_0V^2}{2}, \quad D = \frac{u}{V}S = ut, \quad U = \left(\frac{m_p}{q}\right)\frac{nV^2}{2} = \frac{M_0V^2}{2C}, \tag{16}$$

where $C$ is an positive electric charge of $M_0$.

Typical computations for Earth and interplanetary space vehicles are presented in Fig. 20 - 22 (typical probe, Fig. 20 and Moon spaceship, Figs. 21, 22, - Project 1).

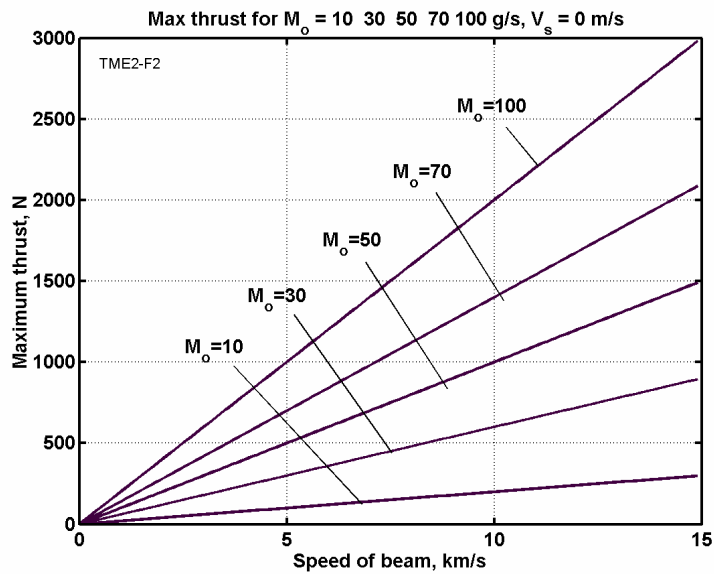

**Fig. 20**. Maximum thrust (drag) versus the particle beam speed and mass flow [g/s] for typical Earth satellite.

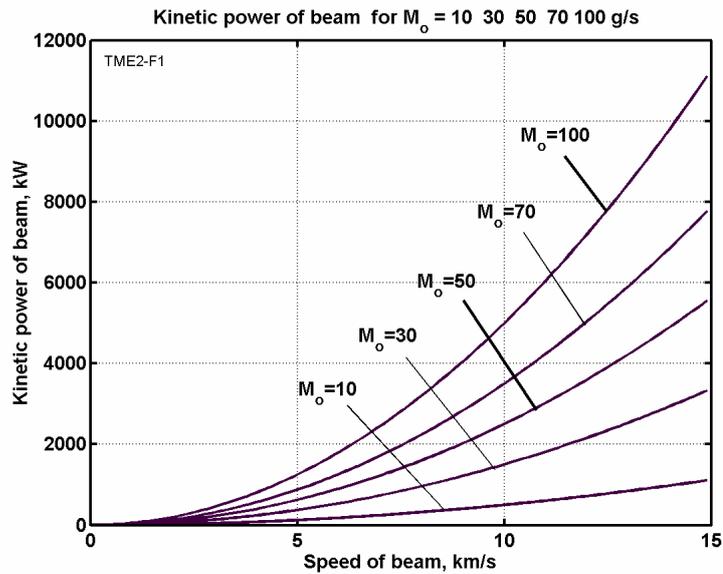

**Fig.21**. Energy is requested for producing of the beam via the beam speed and mass flow for Moon ship.

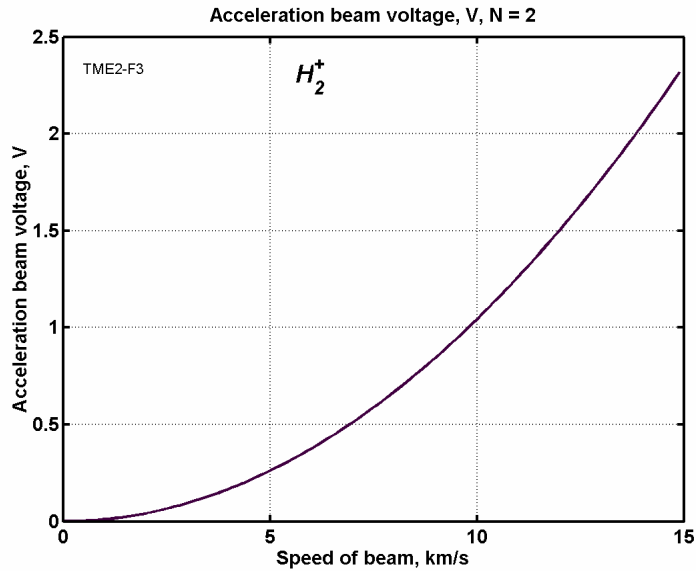

**Fig.22**. Voltage requited for accelerating of beam via a particles speed. The beam is $H_2^+$.

**10. Macro-particle beam and projectile**. The developed theory may be applied to any macro-particle beam or projectiles. An electrostatic gun may be used for its acceleration [27]. The projectile has a charge which is lost after it exits the gun (charged energy will be returned to installation). There is no big problem with flight through the Earth atmosphere for projectiles. The loss energy is about 10 - 20%, the heat is small if projectile has a sharp tip (see [21], [24]). The ship uses a kinetic energy of projectile and projectile matter as described in [29], [30].

The offered method may be applied vareiouly. We consider only application to interplanetary spaceship and interstellar space probe.

# Project 1
*(interplanetary spaceship having weight 3 tons and speed 12 km/s,)*

Assume we want to estimate the parameters of an interplanetary manned spaceship for Moon and Mars having weight of 3 tons and a final speed 12 km/s started from Moon, Mars, or Earth's tower having height of 80 km. We use the theory beam reflection, Figs. 20 - 22 and find the estimation time, thrust, speed, distance. The trained astronauts can stand the overload 8g.

The request power for beam flow 1, 2, 3 kg/s are 312, 625, 937 MW respectively. Power for ionization are 127, 254, 391 MW respectively. That is power of a middle electric station.

# Project 2

*(interstellar probe having speed 30.000 km/s, weight 100 kg)*

Let us assume we want to estimate an interstellar probe which can reach the nearest solar systems. As known they are located about 3 - 4 light years from our Sun. That means the apparatus having speed $\beta_S$ = 0.1c ($V_S$ = 30,000 km/s) can reach them in 30 - 40 years. Remander, "Voyager-1" was flown for 30+ year, sending information up to present time. But it has speed only 20 km/s and was reached only the boundary of Solar system (about 2 billions km).

Assume, the weight of interstellar probe is 100 kg. If distance of acceleration is S = $3 \times 10^{13}$ m (200 AU) the acceleration and acceleration time must be

$$a = \frac{V_S^2}{2S} = \frac{9 \cdot 10^{14}}{2 \cdot 3 \cdot 10^{13}} = 15 \ m/s^2, \quad t = \frac{V_S}{a} = \frac{3 \cdot 10^7}{15} = 2 \cdot 10^6 \ \text{sec} = 23 \ days .$$

The thrust, requested acceleration energy and power are

$$T = aM_S = 15 \cdot 100 = 1500 \ N, \quad W = \frac{M_S V_S^2}{2} = \frac{100 \cdot 9 \cdot 10^{14}}{2} = 4.5 \cdot 10^{16} \ J, \quad P = \frac{W}{t} = \frac{4.5 \cdot 10^{16}}{2 \cdot 10^6} = 22.5 \ GW ,$$

The mass of the beam flow, and energy spent by beam station are (Eq. (3),(5))

$$M_0 = \frac{T\sqrt{1-\beta^2}}{2c(\beta - \beta_S)} = \frac{1500\sqrt{1-0.01}}{2 \cdot 3 \cdot 10^8 (0.1-0.05)} = 5 \cdot 10^{-5} \ kg/s,$$

$$P_B = \frac{M_0 c^2}{2} \frac{\beta^2}{\sqrt{1-\beta^2}} = \frac{5 \cdot 10^{-5} \cdot 9 \cdot 10^{16} 0.01}{2\sqrt{1-0.01}} = 22.5 \ GW$$

Here $\beta_S = 0 \div 0.1$ . We take the average value $\beta_S$ = 0.05 . Notice that $P_B = P$, that means our installation transfer the station energy to ship with efficiency = 1. Unfortunately, this energy is very high. Tens of electric power stations must accelerate this probe in 23 days. We cannot decrease this amount by any methods because that is a minimum energy required by space probe.

Divergence D for u = 0.01 m/s, voltage U (n =1), and plasma temperature $T_p$ are

$$D = \frac{u}{c} \frac{\sqrt{1-\beta^2}}{\beta} S = \frac{0.01}{3 \cdot 10^8} \frac{3 \cdot 10^{13}}{0.1} = 10 \ km, \quad T_p = \frac{mu^2}{ik} = \frac{1.67 \cdot 10^{-27} 10^{-4}}{3 \cdot 1.38 \cdot 10^{-23}} = 0.4 \cdot 10^{-8} \ \ ^0K,$$

$$U = \left(\frac{m_p}{q}\right) \frac{nc^2}{2} \frac{\beta^2}{\sqrt{1-\beta^2}} = \left(\frac{1.67 \cdot 10^{-27}}{1.6 \cdot 10^{-19}}\right) \frac{1 \cdot 9 \cdot 10^{16}}{2} \frac{0.01}{1} = 4.7 \cdot 10^6 \ \ V.$$

The power of dissociations ($H_2 \rightarrow H + H$, 2.2 eV) and ionization ($H \rightarrow H^+$, 13,6 eV) are equal

$$E_i = 1.6 \cdot 10^{-19} (e_d + e_i) \frac{M_0}{m_p n} = 1.6 \cdot 10^{-19} (2.2 + 13.6) \frac{5 \cdot 10^{-5}}{1.67 \cdot 10^{-27} \cdot 1} = 75.7 \ kW .$$

In given case (comparison with $P_B$ above) this value is small and we can negligee it. But into planetary flight ($V \approx 8$ - 30 km/s and large $M_0$) this energy is essential.

# Discussion

In [34] G.A. Landis writes about using particle beams for interstellar flight. The beam is braked by a magnetic sail. Unfortunately, as with most other works in this field, his work also contains only common speculations. No theory, no mathematical models, no computations. More then ten years authors investigate the magnetic sail, but not its theory, no formulas which allows correct calculation or to estimate the magnetic sail drug. Landis offered the beam temperature 45 °K. The theory in this article is shown that this temperature gives the beam divergence which does not allows the interstellar flights. Absolutely unsubstantiated statement that magnetic sail reflects beam in thousands kilometers diameter. The estimations shows for high speed particles especially relativistic particles the affective diameter equals some meters and magnetic field must be powerful. In additional the magnetic sail is impossible at present time: electric ring needs in cryogenic temperature and spaceship must have power cryogenic equipment because the Sun will warm the ring for any heat insulation; for starting the ring needs a power electric station; a special equipment is necessary for displacing the ring of 100 km diameter into space; if the ring temperature exceeds a critical cryogenic temperature in any ring place, the ring explodes. The ring weight is big (22 tons for diameter 100 km), the produced magnetic field is very weak ($10^{-6}$ Tesla). The magnetic sail does not have active control. That means the ship will move in one (non-control) direction and a ship mission will useless. These obvious defects makes impossible the application of the magnetic sail with little or no progress in solution these problems since 1988.

 The suggested method does not require a magnetic sail. That used the electrostatic sail [27] and AB-Ramjet engine offered by author early. This sail is light (100 - 300 kg), cheap, and has tens kilometers (hundreds km for low beam density) of the effective radius. For example, for solar wind the magnetic effective radius decreases proportional $1/R^2$ (where R is distance of the sail from the Sun), electrostatic effective radius decreases approximately $1/R$ (see [27]). That is very important advantage.

## Conclusion

 The offered idea and method use the AB-Ramjet engine suggested by author in 1982 [3, 4, 6, 8, 9, 12, 14 - 16, 18]  and detail developed in [19]. The installation contains an electrostatic particle collector suggested in 1982 and detail developed in [27, 31]. The propulsion-reflected system is light net from thin wire, which can have a large area (tens km) and allows to control thrust and thrust direction without turning of net (Fig.5). This new method uses the ultra-cold full neutral relativistic plasma and having small divergence. The method may be used for acceleration space apparatus (up relativistic speed) for launch and landing Space apparatus to small planets (asteroids, satellites) without atmosphere. For Earth offered method will be efficiency if we built the tower (mast) about $40 \div 80$ km height [20, 25]. At present time that is the most realistic method for relativistic probe.

## *Acknowledgement*

 The author wishes to acknowledge Richard Cathcart for correcting the English and offer useful advices.